\documentclass{aa}
\usepackage{times}
\usepackage{psfig}

\def\lsim{\;\raise0.3ex\hbox{$<$\kern-0.75em\raise-1.1ex\hbox{$\sim$}}\;}
\def\gsim{\;\raise0.3ex\hbox{$>$\kern-0.75em\raise-1.1ex\hbox{$\sim$}}\;}
\def \gr{$\gamma$-ray }

\def\beq{\begin{equation}}
\def\enq{\end{equation}}
\def\begar{\begin{eqnarray}}
\def\endar{\end{eqnarray}}
\def\mathnew{\mathsurround=0pt}
\def\simov#1#2{\lower .5pt\vbox{\baselineskip0pt \lineskip-.5pt
        \ialign{$\mathnew#1\hfil##\hfil$\crcr#2\crcr\sim\crcr}}}

\def\cmc{\rm ~cm^{-3}}

\def\kms{\rm ~km~s^{-1}}

\def\etal{{ et al. }}
\def\bs{{\it BeppoSAX~~}}
\def \sax {{\it BeppoSAX}}
\def \src {IC443}
\begin{document}

\thesaurus{(02.01.1; 02.18.5; 09.03.1; 09.03.2; 09.09.1: \src; 09.19.2)}

\title{Hard X-ray emission from IC443:
evidence for a shocked molecular clump? }

%\\ {\bf  Draft, 9/2/99 } \\
%\end{center}

\author{F. Bocchino\inst{1,2}
\and A. M. Bykov\inst{3}
}
\institute{         
       Astrophysics Division, Space Science Department of ESA, ESTEC,
              Postbus 299, 2200 AG Noordwijk, The Netherlands
\and
       Osservatorio Astronomico di Palermo, Piazza del Parlamento 1, 
       90134 Palermo, Italy
\and
A.F. Ioffe Institute for Physics and Technology,  
St. Petersburg, Russia, 194021}

\offprints{e-mail: byk$@$astro.ioffe.rssi.ru}

\date{Received  / Accepted }

%\maintitlerunninghead{}
\titlerunning{Hard X-rays from IC443}
\maketitle

\begin{abstract}

We report \bs observations of spatially resolved hard X-ray emission
from IC~443, a supernova remnant interacting with a
molecular cloud. The emission is shown to come from two localized
features spatially correlated with bright molecular emission regions. 
Both hard X-ray features have soft X-ray counterparts, in one case
shifted by $\sim 2^\prime$ toward the remnant interior.  The
spectra of X-ray emission from both isolated features have photon index
$\lsim 2$ in the MECS regime.  The emission detected from the remnant
with PDS detector extends up to 100 keV. We discuss the observed
properties of the hard X-ray features in relation to non-thermal emission
from shocked molecular clumps and pulsar wind nebula.


\end{abstract}

\section{Introduction}
Massive stars that are the likely progenitors of core collapsed supernovae
are expected to be spatially correlated with molecular clouds.
IC~443 (G189.1+3.0)  is one of the best space laboratories to study rich
phenomena accompanying
supernova interaction with a molecular cloud. This was established
from observed shock-excited molecular line emission from OH, CO and
H$_2$ (e.g. Burton \etal 1988; van Dishoeck \etal 1993;
Cesarsky \etal 1999). Observations of OH masers and CO clumps towards
IC~443, W28 and W44 also provide convincing
evidence of molecular cloud interaction in this case (Claussen et al 1997).
%Frail \& Mitchell 1998).

IC 443 was a target of  X-ray observations with { \it HEAO 1} (Petre
\etal 1988), {\it Ginga} (Wang \etal 1992), {\it ROSAT} (Asaoka \&
Aschenbach 1994) and {\it ASCA} (Keohane \etal 1997, K97 hereafter).
The soft X-ray 0.2--3.1 keV surface brightness map of IC 443 from the
{\it Einstein} Observatory (Petre \etal  1988) shows bright features in
the northeastern part of the remnant as well as bright soft emission
from the source central part.  The presence of nearly uniform X-ray
emission from the central part of the remnant was clearly seen also by
{\it ROSAT} (Asaoka \& Aschenbach, 1994), and corresponds to the
emission from hot (T$\sim 10^7$ K), low density gas interior to the
shock.  {\it ASCA GIS} observations by K97 discovered the localized
character of the hard  X-ray emission. They concluded that most of the
2-10 keV {\it GIS} photons came from an isolated emitting feature and
from the southeast elongated ridge of hard emission. Preite-Martinez et
al. (1999) have reported a hard ($>14$ keV) component with
BeppoSAX/PDS and two hot spots corresponding to the ASCA sources of
K97 using the MECS.

The supernova remnants (SNRs) that are likely candidates to be
$\gamma$-ray sources in {\it CGRO EGRET} observations (Esposito \etal
1996) also show evidence for interaction with molecular gas.  In the
case of IC 443, the molecular line emission region is partially inside
the {\it EGRET} \gr detection circle.

%Chevalier (1999) discussed the evolution of supernova remnants in
%molecular clouds and concluded that many aspects of the multi-wavelength
%observations could be understood in a model where the remnants evolve
%in the inter-clump medium of a molecular cloud.  

Theoretical models of SNRs interacting with clouds 
were discussed recently by Sturner \etal (1997),
Baring \etal (1999), Chevalier (1999).
The non-thermal multi-wavelength spectrum of a SNR interacting
with a molecular cloud was studied by Bykov \etal (2000). They
showed that the propagation of a radiative shock wave within a molecular
cloud lead to a substantial non-thermal emission both in hard X-rays
and in $\gamma$-ray.  The complex structure of molecular cloud
consisting of dense massive clumps embedded in the inter-clump medium
could result in localized sources of hard X-ray emission correlated
with both bright molecular emission and extended source of non-thermal
radio and \gr emission. In this Letter, we present the archive BeppoSAX
observations of IC443 and, in particularly, we discuss the features
detected in the MECS hard X-ray mosaiced map of the remnant.

%WHAT ABOUT SUBSTITUTING IT WITH SOMETHING LIKE THIS: It is important to
%understand the complex structure of molecular cloud, consisting from
%dense massive clumps embedded in the inter-clump medium, and the regions
%showing non-thermal X-ray emission inside SNRs. In fact, it is not
%clear to which extent the non-thermal X-ray emission from SNRs may be
%due to acceleration processes located into the shock (typically
%synchrotron) or into high density clumps of molecular clouds (typical
%bremssthralung). To accomplish this task, it is crucial to investigate
%the morphology and spectral characteristic of compact hard X-ray sources
%inside SNRs and their correlation with bright molecular emission and
%the extended source of non-thermal radio and \gr emission.  In this
%Letter, we report on BeppoSAX observations of two compact hard X-ray
%sources inside IC443.

%The observed structure of IC~443
%seems consistent with a highly inhomogeneous remnant model in which the
%hard emission is localized at the southeastern part
%of the remnant.

\section{Observations and data analysis}

Results from the the Medium-Energy Concentrator
Spectrometer (MECS; 1.8--10~keV; Boella et al. 1997)
and the Phoswich Detection System (PDS; 15--300~keV; Frontera
et al. 1997) on-board \sax\ are presented. 

\begin{figure*}
  \centerline{\hbox{
    \psfig{figure=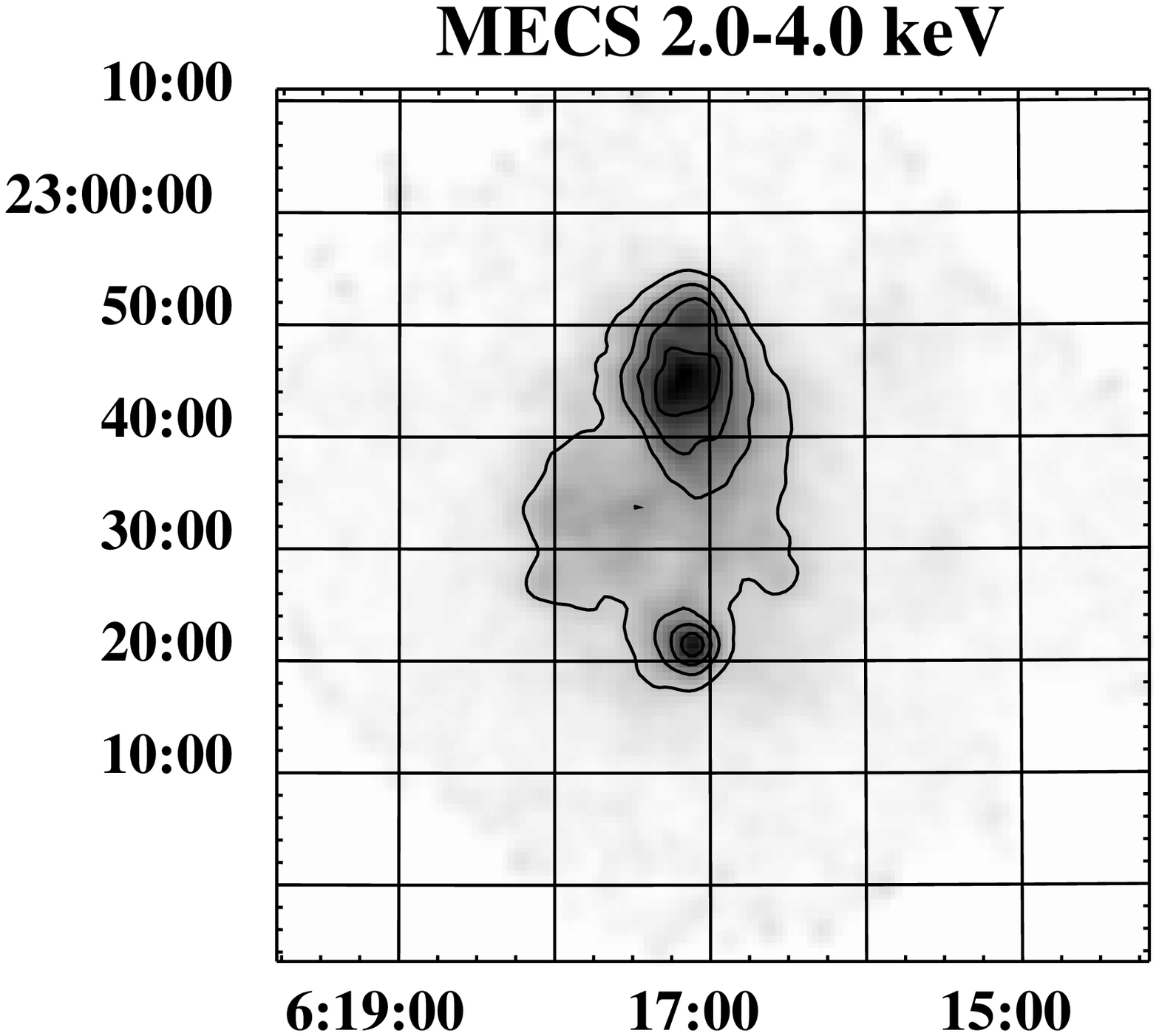,width=8.5cm}
    \psfig{figure=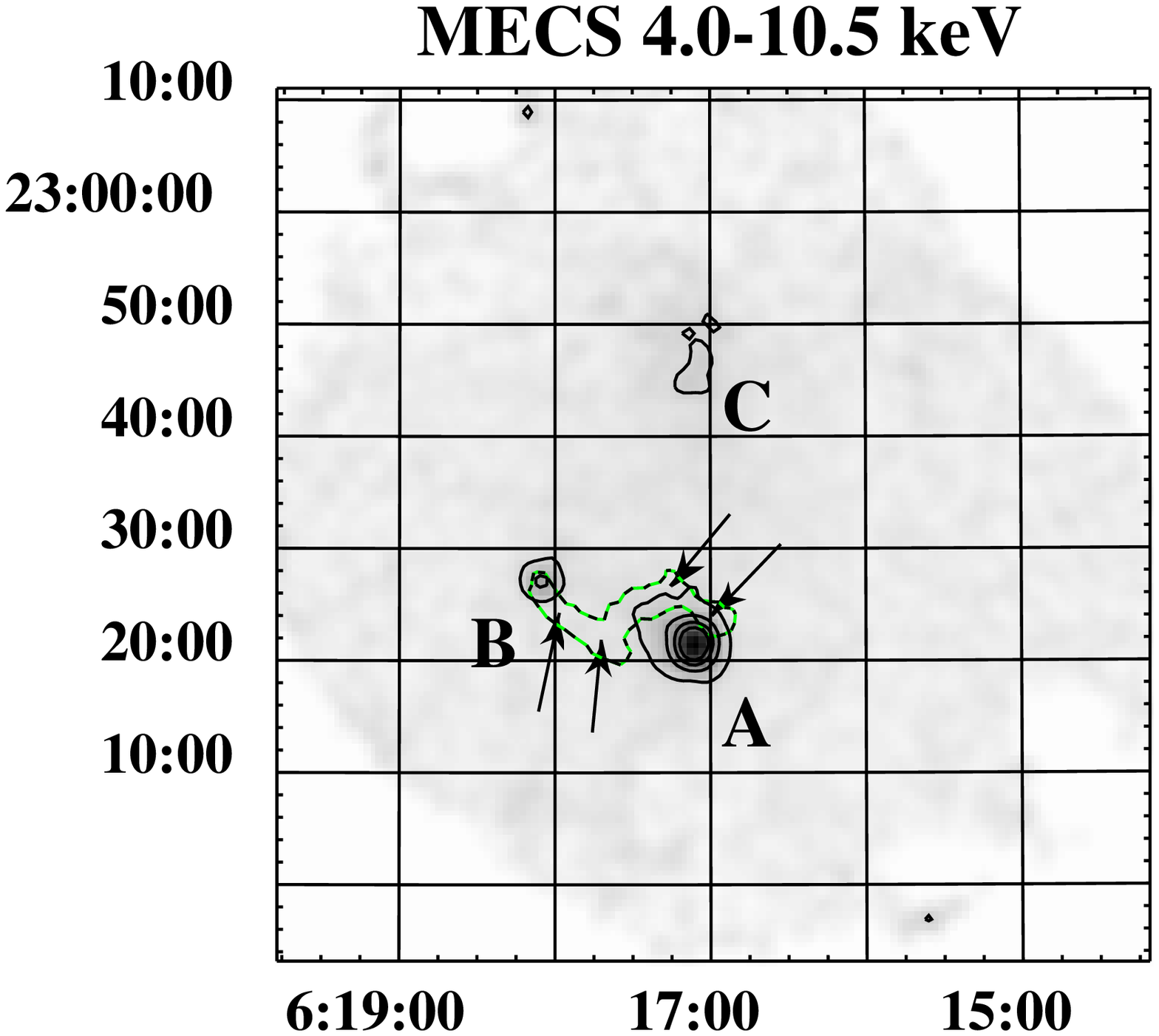,width=8.5cm}
  }}
  \caption{MECS vignetted and exposure corrected mosaic of the four
  IC443 observations listed in Table \protect\ref{obs}. Grayscale is
  linear and contours are 20\%, 40\%, 60\% and 80\% of the peak (3.6
  and $2.2\times 10^{-3}$ cnt s$^{-1}$ arcmin$^{-1}$ for the soft and
  hard image, respectively).  In the right panel, the sources discussed
  in the text are marked A, B and C, and dashed contour correspond to a
  $v=1-0S(1)$ molecular H line intensity of $6.6\times 10^{-14}$ erg
  cm$^{-2}$ s$^{-1}$ reported by Burton et al. (1988); arrows indicate
  the location of peaks with an H line intensity at least 14 times the
  contour levels.}

  \label{mecs}
\end{figure*}

%\begin{figure}
%  \centerline{\hbox{
%    \psfig{figure=rate_le_bl4_0.12_2.0_imh_smo.ps,width=8cm}
%  }}
%  \caption{LECS vignetted and exposure corrected mosaic of IC443.
%  Grayscale is linear and contours are 20\%, 40\%, 60\% and 80\% of the
%  peak ($5.0\times 10^{-3}$ cnt s$^{-1}$ arcmin$^{-1}$).}
%
%  \label{lecs}
%\end{figure}

%All these instruments are
%coaligned and collectively referred to as the Narrow Field Instruments,
%or NFI.  The MECS consists of two grazing incidence telescopes with
%imaging gas scintillation proportional counters in their focal planes.
%The LECS uses an identical concentrator system as the MECS, but
%utilizes an ultra-thin entrance window and a driftless configuration to
%extend the low-energy response to 0.1~keV.  The non-imaging HPGSPC
%consists of a single unit with a collimator that was alternatively
%rocked on- and 180\arcmin\ off-source every 96~s during the
%observation.  The non-imaging PDS consists of four independent units
%arranged in pairs each having a separate collimator. Each collimator
%was alternatively rocked on- and 210\arcmin\ off-source every 96~s
%during the observation.

\begin{table}
\caption{Public BeppoSAX Observations of \src}
\label{obs}
\medskip
\centering\begin{minipage}{8.7cm}
\begin{tabular}{lccc} \hline
Obs. & T$_{exp}^{MECS}$ & Coord (J2000) & Date \\ 
       & ksec &               \\ \hline 

IC443H & 21.5 & $6^h~17^m~05^s$ $+22^d~22^m~19^s$ & 11/4/1999 \\
IC443E & 48.4 & $6^h~18^m~00^s$ $+22^d~34^m~01^s$ & 11/4/1998 \\
IC443C & 67.1 & $6^h~16^m~01^s$ $+22^d~28^m~55^s$ & 18/10/1997 \\
IC443N & 23.8 & $6^h~17^m~12^s$ $+22^d~48^m~50^s$ & 20/10/1997 \\
\hline
\end{tabular}
\end{minipage}
\end{table}

We have used all the 4 observation of IC443 available in the public
\sax\ archive; Table \ref{obs} lists their coordinates and exposure
times.  Good data were selected from intervals when the elevation angle
above the Earth's limb was $>$$4^{\circ}$ and when the instrument
configurations were nominal, using the SAXDAS 2.0.0 data analysis
package.

%The soft LECS image is dominated by the thermal emission of the
%remnant, which was earlier reported by Petre \etal (1988), Asaoka \&
%Aschenbach (1994) and K97.  The 2.0-4.0 MECS images show an emission

Fig. \ref{mecs} shows the MECS mosaics of IC443 in different bands. The
images are vignetting and exposure corrected.  The 2.0-4.0 MECS images
show an emission morphology similar to the GIS broad-band maps
presented by K97.  On the other hand, the hard MECS image above 4 keV
is very different. There is weak diffuse extended emission in the north
of the remnant, and two strong compact sources, Src A and B. Table
\ref{srcs} reviews the properties of the two sources. In Fig.
\ref{mecs}, we also marked the weak extended hard X-ray emission with
C. Both the Src A and B were observed by K97; Src A corresponds to the
hard source named ``HXF", while Src B corresponds to the hard extended
emission called ``The Ridge" (also reobserved by Olbert \etal 2000, and
detected by Preite-Martinez et al. 1999 as 1SAX J0618.1+2227). In the
hard X-ray image in Fig. \ref{mecs} we have also reported the contour
of the molecular hydrogen line emissivity observed by Burton et al.
(1988), which shows that the hard X-ray sources are closely correlated
with it.

%Moreover, Src B is at only
%$3^\prime$ from IC443B, a cloud with an observed anhancement of HCO$^+$
%attributed to an enhanced cosmic-ray energy density (Shuter \etal 1986)

\begin{figure*}
  \centerline{\hbox{
    \psfig{figure=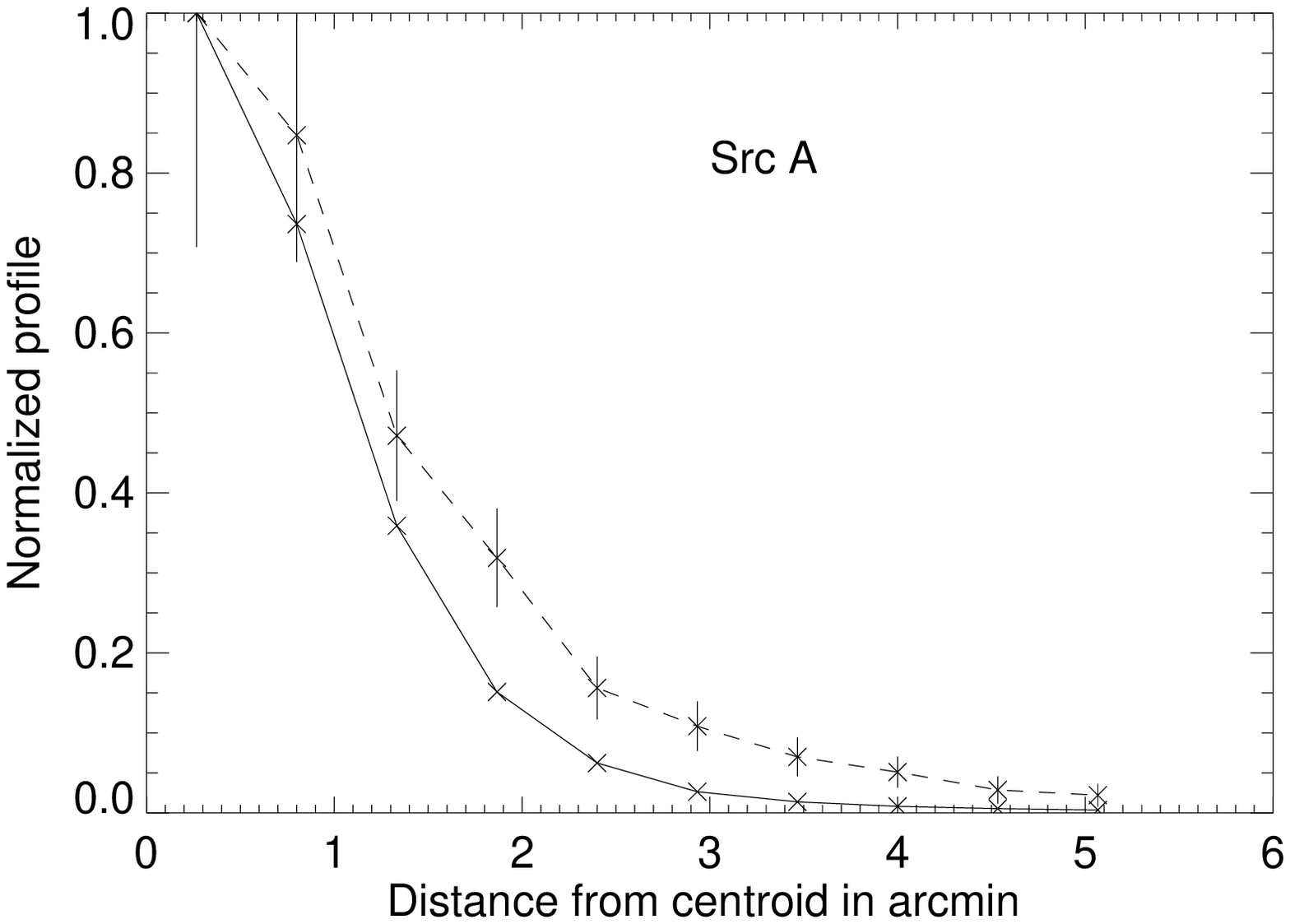,width=8cm}
    \psfig{figure=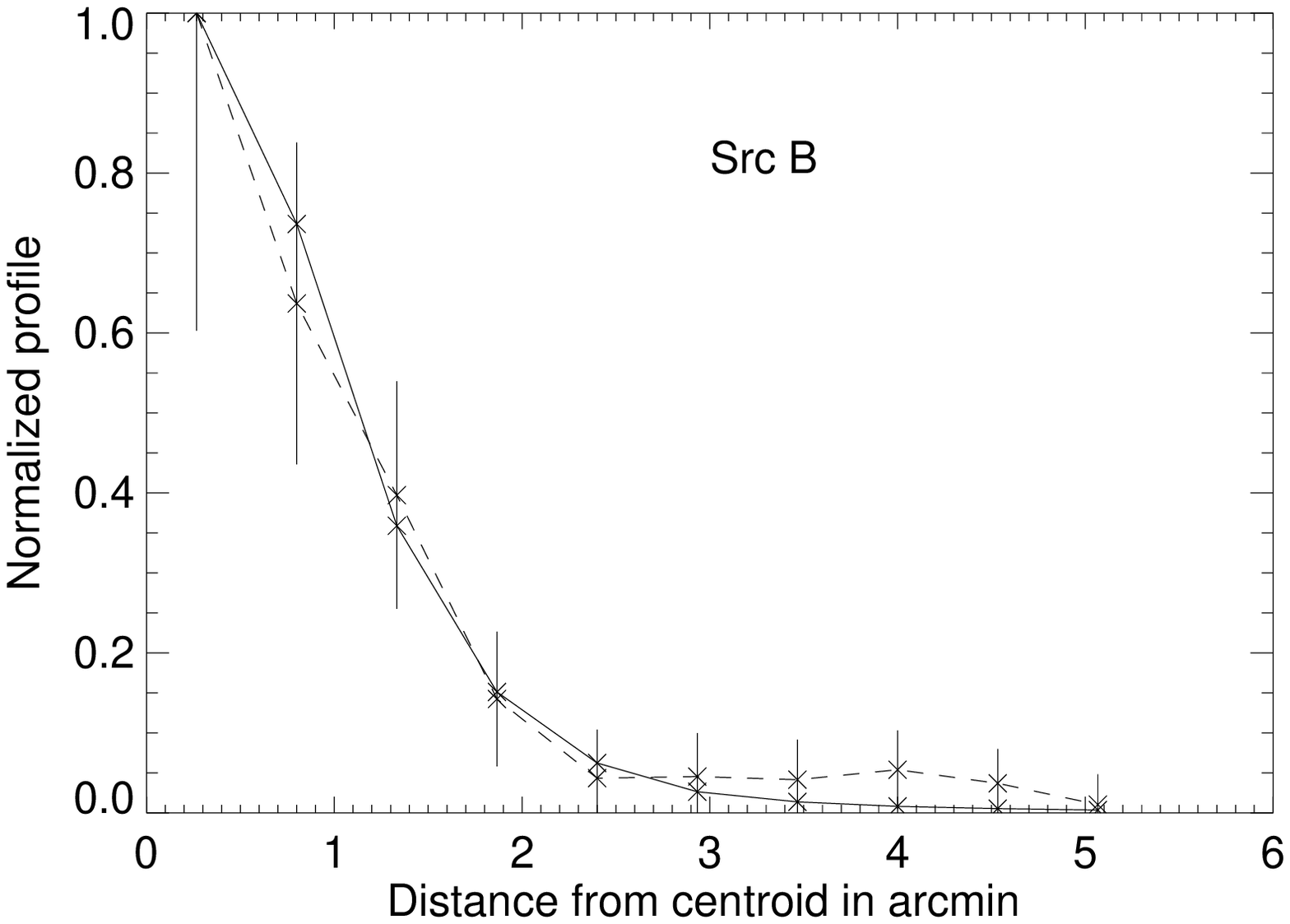,width=8cm}
  }}
  \caption{Hard X-ray profile (4.0-10.5 keV) of the MECS Src A and B
  (dashed line) compared with the expected profile of point-like source
  (solid line) derived from a long MECS observation of the X-ray binary
  Cyg X-1.}

  \label{prof}
\end{figure*}

\begin{table}
\caption{Compact sources detected in the hard MECS images.}
\label{srcs}
\medskip
\centering\begin{minipage}{8.0cm}
\begin{tabular}{lccc} \hline
Name & Count-rate\footnote{In the 4.0-10.0 keV energy range.} & 
Position \\ % & Notes \\ 
       & (s$^{-1}$) & (J2000)       \\ \hline 

Src A & $3.7\pm0.2\times 10^{-2}$ & $6^h~17^m~6.3^s$ $+22^d~21^m~37^s$ \\ % & $22\sigma$ \\ 
%1WGA J0617.1+2221 & $2.8\pm0.3\times 10^{-2}$ & $6^h~17^m~06.1^s$ $+22^d~21^m~30^s$ & $\pm 13^{\prime\prime}$, $7^{\prime\prime}$ from Src A \\ \hline
Src B & $5.9\pm0.6\times 10^{-3}$ & $6^h~18^m~2.7^s$ $+22^d~27^m~28^s$ \\ % & $9\sigma$ \\
\hline

\end{tabular}
\end{minipage}
\end{table}

In Fig. \ref{prof} we report the 4.0-10.5 keV profile of Src A and Src
B compared with the MECS PSF profile in the same energy range. The PSF
profile was derived from a BeppoSAX observation of the very bright
point source Cyg X-1. Src A seems to be extended, while Src B is
consistent with the expected PSF. The latter was reported as an
elongated ridge by K97, but we note that the ASCA GIS pointing
direction is such that the location of Src B is seen at large off-axis
angle, which can account for most of source extension observed by ASCA.
Instead, the BeppoSAX Src A and B are placed  only at
$5^\prime$ and $1^\prime$ off-axis, where the instrumental distorsion
is at its minimum.

\begin{table}
\caption{Summary of MECS spectral fitting results.}
\label{bestfit}
\medskip
\centering\begin{minipage}{7.0cm}
\begin{tabular}{lccc} \hline
Model & kT & $\gamma$ & $\chi^2/dof$ \\
     & keV  &  \\ \hline 
 & \multicolumn{3}{c}{Src A} \\
P.L. only & - & $1.96^{+0.21}_{-0.12}$ & 34/41 \\ \hline
 & \multicolumn{3}{c}{Src B} \\
%%%% P.L. only & - & $3.2^{+0.15}_{0.14}$ & 61/35 \\
P.L.+{\sc mekal} & $1.07^{+0.23}_{-0.31}$ & $1.5^{+0.9}_{-0.9}$ & 41/33 \\
\hline
\end{tabular}
\end{minipage}
\end{table}

We have also investigated if Src A and Src B have soft X-ray
counterpart using archive ROSAT observations. Fig. \ref{pspc} is a PSPC
0.2-2.0 keV image of the region containing MECS Src A and B, with the
MECS 4.0-10.5 keV contours of Fig. \ref{mecs} (right panel)
overimposed.  Src A has an HRI counterpart quoted by K97 and the PSPC
WGACAT source 1WGA J0617.1+2221 with a rate of $2.8\pm0.3\times
10^{-2}$ cnt s$^{-1}$ is located inside the BeppoSAX error circle. Src
B has a weaker PSPC counterpart indicated by the arrow in Fig.
\ref{pspc}, located $\sim 1.5^\prime$ north of the MECS source centroid
and not reported in other PSPC catalogs. We have
verified that the extended hard X-ray emission we have labelled with C
in Fig \ref{mecs} has a thermal origin and it represents the weak
hard tail of the bright soft X-ray nebula visible in the PSPC image.

We have performed spectral fits of the MECS spectra of Src A and B
using the {\sc mekal} optically thin thermal plasma model modified by
the interstellar absorption assuming the standard abundances.  Spectra
were extracted from circular regions with radii of $4^\prime$ and
$3^\prime$ for Src A and B, respectively. We have used the IC443H
observation for Src A and IC443E observation for Src B. Spectra
have been rebinned to have at least 20 counts per channel. The MECS
background was collected in an annulus between $5^\prime$ and
$8^\prime$ for Src A, while we have used the standard background for
Src B because no source-free background region could be found around it.

% which is immersed in the thermal emission and a proper background
%region could not be found around it.

A thermal origin for the spectrum of Src A is strongly rejected
($\chi^2/d.o.f.=111/41$), while a power law model nicely fit the data
(Table \ref{bestfit}). 
The spectrum of Src B cannot be reproduced by a power law or by a
thermal emission model alone ($\chi^2/d.o.f. =$ 61/33 and 78/33
respectively) but requires a combination of the two, reported in Table
\ref{bestfit}.  

\begin{figure}
  \centerline{\hbox{
    \psfig{figure=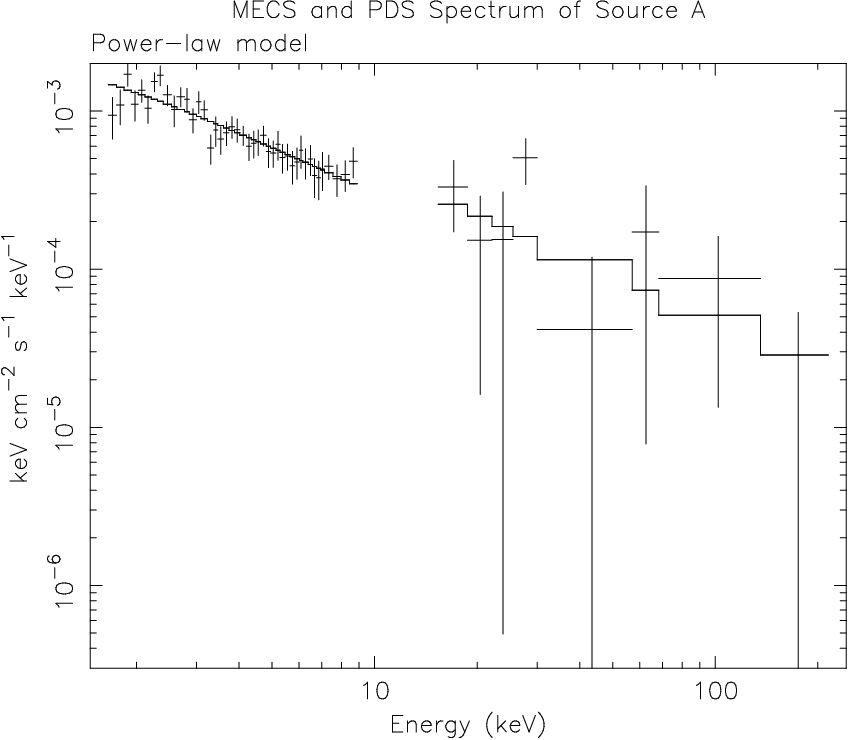,width=8cm,angle=-90}
  }}
  \caption{MECS and PDS spectrum of Src A. The power law model which
  best fit the data (Table \protect\ref{bestfit}) is also reported with
  residuals.}

  \label{srca_eufspec}
\end{figure}

We have also analyzed the PDS spectra. Background was subtracted using
the OFF collimator positions; we have verified that no other
contaminating sources are present in the background positions. Spectra
were collected using the Variable Rise Time Threshold mode and have
been rebinned following a logarithmic scheme suggested by the PDS
hardware group to not undersample the instrument spectral resolution.
Table \ref{rates} reports the observed countrates and flux in different
energy ranges. A joint fit of Src A MECS spectrum and IC443H PDS
spectrum (shown in Fig. \ref{srca_eufspec}) yields best-fit
results similar to the ones reported in Table \ref{bestfit} and a
PDS/MECS normalization ratio of $1.40\pm0.45$. This value is consistent
with the expected value of 1, also considering that the expected
constribution of Src B in the PDS spectrum of IC443 could be
up to $\sim 50\%$.  As for Src B, the joint fit yields a PDS/MECS
ratio of $2.8\pm2.5$, we argued that the PDS spectra of IC443E is
heavily contaminated by the brighter Src A. In fact, using the best-fit
model of Src A and considering the triangular spatial response of the
PDS collimator, we predict a Src A contribution of $\sim 6\times
10^{-2}$ cnt s$^{-1}$ in the PDS spectra of Src B between 15 and 30
keV, consistent with the observed rate (Table \ref{rates}). Therefore,
a proper analysis of the PDS spectra of Src B cannot be done.

\begin{table}
\caption{Observed background subtracted count-rates and 
unabsorbed fluxes in different energy 
ranges, MECS for $E<10$ keV and PDS for $E>10$ keV.}
\label{rates}
\medskip
\centering\begin{minipage}{5.8cm}
\begin{tabular}{lrr} \hline
Band & Count-rate & Flux $\times 10^{-12}$ (\footnote{Computed 
using the best-fit models and 
values reported in Table \ref{bestfit}.}) \\ 
keV  & ($10^{-2}$ s$^{-1}$) & erg cm$^{-2}$ s$^{-1}$  \\ \hline 
 & \multicolumn{2}{c}{Src A} \\
 2--10 & $8.2\pm0.2$ & $7.5 \pm 1.2$  \\
 15--30 & $11.5\pm 3.0$ & $9.1 \pm 2.4$ \\
 30--100 & $7.2\pm4.3$ & $11.1 \pm 6.6$ \\
 100-220 &  $5.8\pm3.7$ & $41.5 \pm 26.7$ \\ \hline
 & \multicolumn{2}{c}{Src B} \\
 2--10 & $2.0\pm0.1$ & $1.8 \pm 0.2$\footnote{47\% {\sc mekal}, 53\% P.L.} \\
 15--30 & $4.3\pm2.4$ & $3.4 \pm 1.9$ \\
 30--100 & $<3.9$ & $<1.1$ \\
% 100-220 & too noisy \\
\hline

\end{tabular}
\end{minipage}
\end{table}

\section{Discussion}

%Smaller scale clumps down to a scale 1" were
%also observed by Richter \etal (1995; COMM: THIS REFERS TO IC443?). 

\begin{figure}
  \centerline{\hbox{
    \psfig{figure=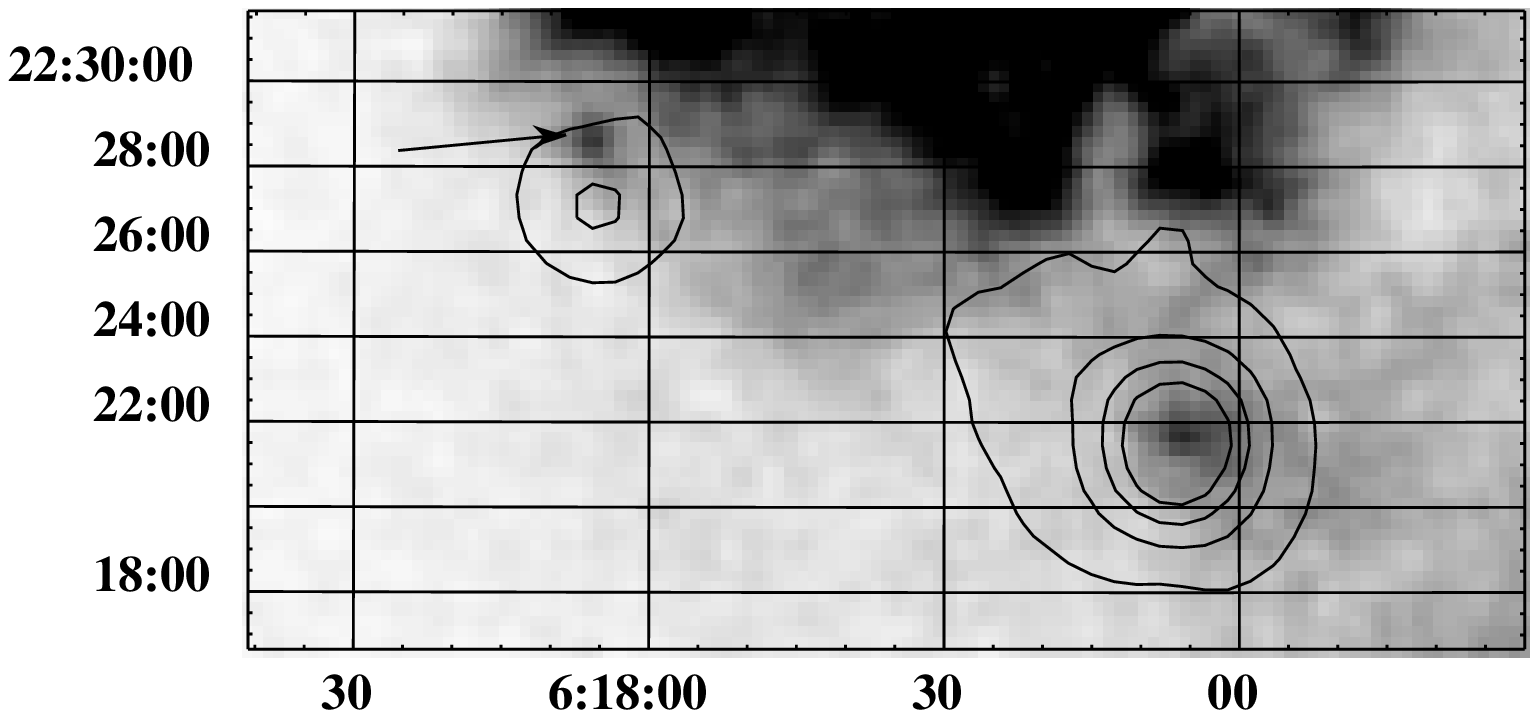,width=8cm}
  }}
  \caption{0.2-2.0 keV PSPC image of the region cotaining the two MECS
  hard X-ray sources. The hard X-ray contours of Fig.
  \protect\ref{mecs} are overlaid. The arrow shows the
  soft X-ray counterpart of Src B.}

  \label{pspc}
\end{figure}

The compact hard X-ray features may be interpreted in
terms of the interaction between the remnant shock and a molecular
cloud.  Bykov \etal (2000) have presented a model of non-thermal
emission from an evolved SNR
interacting with a molecular cloud, but
the predicted diffuse hard X-ray
emission in the MECS bandwidth is
below the sensitivity threshold.
On the other hand, several clumps have been observed in the
molecular cloud near IC443.  The spatial scales of the 
molecular clumps 
from IR line observations are in the range from 150$^{\prime \prime}$
down to 1$^{\prime \prime}$.  The non-thermal X-rays from clumps were
modelled by Bykov \etal (2000), who predicted a hard spectrum with a
photon index below 2.0. The emission is originated from electrons of
energy below 1 GeV.
The predicted flux density is $\sim 6\times
10^{-5}$ keV cm$^{-2}$ s$^{-1}$ keV$^{-1}$
at 10 keV for a 30 $\kms$ shock in a clump of
half parsec radius and density of 10$^4~ \cmc$.  Using a power law with
$\gamma=1.5$, this corresponds to $\sim 3\times 10^{-2}$
MECS cnt s$^{-1}$ in the 4.0--10.5 keV energy range and $\sim
10^{-2}$ PDS cnt s$^{-1}$ in the 15--30 keV. The predicted count-rates
are consistent with the observed rate of Src B (Table \ref{rates}),
having in mind that $10^4$ cm$^{-3}$ density is a representative number
and other parameters could easily account for the difference.  Because
of hard spectrum and heavy absorption, localized spots of few arcmin
size would be seen only in hard X-rays. Thus, the compact Src B
correlated with bright spot of molecular hydrogen emission (Fig.
\ref{mecs}) can be shocked molecular clump. The observed soft X-ray
brightening having apparent shift from hard X-ray features towards the
SNR interior (Fig. \ref{pspc}), can be attributed to the effects of the
shocked clump edges. The morphology and the spectrum of the Src A are
somewhat different from that of Src B.

As already discussed by K97, a low-luminosity pulsar nebula is a
plausible explanation for isolated hard X-ray sources (especially for
the slightly extended Src A), but we do not expect to see two of them
in one SNR\footnote{However, Asaoka \& Aschenbach (1994) pointed out
that the soft X-ray emission is consistent with the presence of two
large SNRs, $10^3$ and $10^5$ yr old.}. High resolution Chandra or XMM
data are needed to distinguish between pulsar nebula and shocked clump
interpretation.  We have also performed a timing analysis of MECS
events collected inside a circle of $3^\prime$ radius in the 4.0--10.5
energy range for both Src A and B, and we found no pulsations at 99\%
confidence level, with an upper limit of the pulsed fraction of a
sinusoidal signal in the $10^{-3}-64$ Hz frequency range of 5\% and
6\%, respectively, more stringent than the one derived by K97 using
ROSAT data.

We have also estimated the Src A and B EGRET \gr $E>100$ MeV flux
contribution, using the best-fit models reported in Table
\ref{bestfit}, and they are $3.0^{+0.5}_{-0.4}\times 10^{-8}$ and
$100^{+380}_{-92}\times 10^{-8}$ photons cm$^{-2}$ s$^{-1}$. The
expected \gr flux of Src B therefore is compatible with the value of
$50.0\pm 3.0\times 10^{-8}$ photons cm$^{-2}$ s$^{-1}$ observed by
Esposito et al.  (1996). Gamma-ray observations with forthcoming {\it INTEGRAL} 
mission with expecting angular resolution about 12$^\prime$ could better
constrain the spectrum of the sources above 30 keV.

\begin{acknowledgements}
The authors thank M. Orlandini for his help in analyzing the 
PDS data and the referee (J.Keohane) for very constructive comments.
The work of A.M.B was supported by the INTAS 96-0390 and
INTAS-ESA 99-1627 grants. F.B. acknowledges an ESA Research 
Fellowship.

\end{acknowledgements}
\appendix

\section*{References}

\begin{flushleft}
\medskip

\small

%Anders E., Grevesse N., 1989, Geocosmochimica Acta 53, 197

Asaoka I., \& Aschenbach B. 1994, A\&A, 284, 573

Boella G., Chiappetti L., Conti G., et al., 1997, A\&AS 122, 327

Baring M.G., Ellison D.C., Reynolds S.P., et al., 1999, ApJ, 513, 311  

Burton M.G., Geballe T., Brand P., et al., 1988, MNRAS,  231, 617

Bykov A., Chevalier R., Ellison D., Uvarov, Yu., 2000, ApJ, 538, 203

Cesarsky D., Cox P., Pineau d. Forets G., et al., 1999, A\&A, 348, 945

Chevalier, R.A. 1999,  ApJ,  511, 798

Claussen M.J., Frail D.A., Goss W.M., et al., 1997,
 ApJ, 489, 143

% DeNoyer L.K. 1979, ApJ,  232, L165

Esposito J.A., Hunter S.D., Kanbach G., et al., 1996,  ApJ, 461, 820

%Fesen R.A., \& Kirshner R.P. 1980,  ApJ, 242, 1023

%Frail, D.A., \& Mitchell, G.F. 1998, ApJ, 508, 690

Frontera F., Costa E., Dal Fiume D., et al., 1997, A\&AS 122, 371

% Green D.A. 1986,  MNRAS, 221, 473

Keohane J.W., Petre R., Gotthelf, et al., 1997, ApJ, 484, 350

%Manzo G., Guarrusso S., Santangelo A., et al., 1997, A\&AS 122, 341

% Mewe R., Gronenschild E., van den Oord G. 1985, A\&AS, 62, 197

%Morisson D., McCammon D., 1983, ApJ 270, 119

%Mufson, S.L., McCollough, M.L.,  Dickel, J.R., Petre, R.,
%White, R., \& Chevalier, R. 1986, AJ, 92, 1349

Olbert C.M., Clearfield C.R., Williams N.E., Keohane J.W.,
\& Frail D.A., 2000 (in preparation)

% Parmar A.N., Martin D., Bavdaz M., et al., 1997, A\&AS 122, 309

Petre R., Szymkowiak A.E., Seward F.D., et al., 1988,
ApJ, 335, 215

Preite-Martinez A., et al. 1999, AIP Conf. Proc. 510, 73

%Richter, M.J., Graham, J.R., \& Wright, G.S. 1995, ApJ,  454, 277

%Rho, J., \& Petre, R. 1998,  ApJ, 503, L167

% Shuter W.L.H., Williams D., Kulkarni S., et al., 1986, ApJ, 306, 255

Sturner S.J., Skibo J., Dermer, C. et. al. 1997, ApJ, 490, 617  

van Dishoeck E.F., Jansen D., \& Phillips T.  1993, A\&A 279, 541

Wang Z.R., Asaoka I., Hayakawa S., et al., 1992, PASJ,  44, 303

\end{flushleft}
\end{document}